\documentclass[journal]{IEEEtran}   
\usepackage{subfigure}
\usepackage[dvipdfmx]{graphicx,xcolor}
\usepackage{epstopdf}
\usepackage{epsfig}	
\usepackage{cite,amsmath,amssymb}
\usepackage{algorithm}
\usepackage{algpseudocode}
\usepackage{url}
\usepackage{fancyhdr}
\usepackage{mdwmath}
\usepackage{mdwtab}
\usepackage{caption}
\usepackage{amsthm}
\usepackage{lipsum}

\begin{document}

\title{Multi-Agent Deep Reinforcement Learning Based Resource Management in SWIPT Enabled Cellular Networks with H2H/M2M Co-Existence}

\author{Xuehua~Li, Xing~Wei, Shuo~Chen, Lixin~Sun \thanks{Xuehua Li, Xing Wei and Shuo Chen are with the Key Laboratory of Modern Measurement and Control Technology, Ministry of Education, Beijing Information Science and Technology University, Beijing 100101, China (e-mail: lixuehua@bistu.edu.cn, weixing@bistu.edu.cn, chenshuo@bistu.edu.cn).} \thanks{Lixin Sun is with the Baicells Technologies Co., Ltd, Beijing 100094, China (e-mail: sunlixin@baicells.com)}\thanks{This work was supported in part by the National Natural Science Foundation of China under Grant 61901043; the Qin Xin Talents Cultivation Program, Beijing Information Science and Technology University under Grant QXTCP B202101; R\&D Program of Beijing Municipal Education Commission (KM202211232010).}}

\maketitle

\begin{abstract}
    Machine-to-Machine (M2M) communication is crucial in developing Internet of Things (IoT). As it is well known that cellular networks have been considered as the primary infrastructure for M2M communications, there are several key issues to be addressed in order to deploy M2M communications over cellular networks. Notably, the rapid growth of M2M traffic dramatically increases energy consumption, as well as degrades the performance of existing Human-to-Human (H2H) traffic. Sustainable operation technology and resource management are efficacious ways for solving these issues. In this paper, we investigate a resource management problem in cellular networks with H2H/M2M coexistence. First, considering the energy-constrained nature of machine type communication devices (MTCDs), we propose a novel network model enabled by simultaneous wireless information and power transfer (SWIPT), which empowers MTCDs with the ability to simultaneously perform energy harvesting (EH) and information decoding. Given the diverse characteristics of IoT devices, we subdivide MTCDs into critical and tolerable types, further formulating the resource management problem as an energy efficiency (EE) maximization problem under divers Quality-of-Service (QoS) constraints. Then, we develop a multi-agent deep reinforcement learning (DRL) based scheme to solve this problem. It provides optimal spectrum, transmit power and power splitting (PS) ratio allocation policies, along with efficient model training under designed behaviour-tracking based state space and common reward function. Finally, we verify that with a reasonable training mechanism, multiple M2M agents successfully work cooperatively in a distributed way, resulting in network performance that outperforms other intelligence approaches in terms of convergence speed and meeting the EE and QoS requirements.
\end{abstract}
\section{INTRODUCTION}
Accompanied by the emergence of IoT, human society is gradually becoming digitalized and data-driven, which is achieved through key verticals such as connected industries, intelligent transport systems and smart cities. Significantly, machine type communication (MTC) is the main driver of this digitalization, including its massive and critical aspects, as well as near-instant wireless connectivity without direct human intervention. It is envisaged that the number of connected devices will reach 500 billion by 2030 \cite{analytics2018electric}, serving a wide variety of use cases with highly diverse requirements. This evolving paradigm is known as M2M communications. As the number of connected machines grows exponentially, MTC will be a key driver in the growing trend towards making anything that moves autonomous and intelligent.

Unlike conventional H2H communications, M2M applications have their own requirements in terms of EE, periodic data transmission and strict delay limitation. In order to deploy M2M communications, low power wide area networks (LPWAN) have been specifically designed for MTC, which are usually extremely power efficient and are able to provide connectivity over a long range \cite{huda2019six}. However, the date rate and supported use cases of LPWAN are too limited to sustain some MTC applications. Cellular networks have been considered the key enabler for M2M communications due to the diverse services access and rates they offer through more extensive networks and larger bandwidths. However, as the supported H2H devices and the network itself are usually too power-hungry and costly for many energy-constrained MTC applications, EE optimization is an urging problem for cellular networks with H2H/M2M coexistence.

SWIPT has been produced as a promising technology to enhance the EE of the network. Although traditional EH technology can enable mobile devices or base station (BS) to obtain energy from natural environments \cite{mitcheson2008energy}, such as solar and wind, the efficiency of EH is severely affected by the location factors and weather conditions. As a radio frequency (RF)-based EH a technology, in contrast, SWIPT enabled cellular networks with H2H/M2M coexistence allow MTCDs to obtain relatively controllable and stable energy, e.g., macro BS and Wi-Fi access points \cite{gui2018deep,wang2019ul,wang2019data}. In general, it can convert the interference signals to electricity, that is, strong interference increases the amount of EH, but decreases the system throughput \cite{muy2021energy}.

In addition to improving the EE, SWIPT enabled cellular networks with H2H/M2M coexistence are encountering many challenges to satisfy various QoS demands, such as intra-tier and cross-tier interference control, spectrum assignment, the tradeoff between system spectrum efficiency and EH, etc. Consequently, while the SWIPT becomes an available approach for low-cost and pollution-free power supply of a variety of IoT devices \cite{xu2019optimal}, many problems still remain to be solved. To overcome these problems, resource management is one of the key solutions.

Resource management can provide QoS assurance for IoT devices under the inherent scarcity of radio resources. However, as network architecture have developed, diversity and complexity in resource allocation tasks have emerged, resulting in a poor effectiveness of traditional resource management approaches. Fortunately, the RL approaches have been exploited for resource management in recent years because their capability of solving complex problems in a dynamic and uncertain environments \cite{sutton2018reinforcement}. To further enhance the capability of RL with regard to learning from high-dimensional raw input data and making intelligent decisions, a combined technology called based on RL and deep learning--DRL--has shown its great potential. In particular, deep Q networks (DQN), which consist of two phases--an offline deep neural network (DNN) construction phase and an online Q-learning phase--has gained a reputation for handling complicated control problems.
\subsection{Related Work}
Design of energy-efficient resource allocation is essential since most of the MTCDs in IoT are battery-powered with limited battery capacity and charging facilities. In \cite{aijaz2014energy}, the authors have formulated the energy-efficient resource allocation problem as a maximization of effective capacity-based bits-per-joule capacity under statistical delay constrains. Authors in \cite{dawaliby2018joint} have tackled the challenges of scheduling M2M traffic in cellular networks and proposed a cross-layer resource allocation scheme that minimized the energy consumption of the Machine-Type Communication Devices (MTCDs). A realistic energy consumption model for MTC and network battery-lifetime model was developed in \cite{azari2017network}, analytic expressions are derived to demonstrate the impact of scheduling on both the individual and network battery lifetimes. A waiting-line based resource allocation framework was proposed in \cite{shih2016wait}, which guarantees resources for H2H communications while meeting the needs of M2M communications on a first-come first-served basis. Although the aforementioned schemes have improved the EE of cellular networks with H2H/M2M coexistence to some extent, the sustainable operation technology which can significantly enhance EE performance was ignored, such as SWIPT.

By integrating SWIPT with MTCDs, which are able to harvest electromagnetic energy from RF signals, the EE of system can be markedly improved. The very ambient RF harvesting prototype has been presented in \cite{pinuela2013ambient}, in which different energy sources (i.e., RF, thermal and vibration) were compared to prove the technological advantages of RF in terms of power density. In \cite{shih2016two}, authors proposed three different spectrum access schemes for EH-M2M communication with the aims of improving the performance in terms of throughput, delay and energy efficiency. A large-scale M2M network architecture was developed in \cite{huang2017performance}, which incorporates EH and social-aware relays, where the relay is powered by harvesting RF energy and adopts the SWIPT strategy. Authors in \cite{zhou2019energy} proposed a two-stage 3-D matching algorithm to maximize the EE of the transmitter in M2M communication. However, the co-channel interference caused by spectrum sharing and various MTC traffic types have not been taken into consideration in the above-mentioned researches. In addition, the high computation complexity of energy consumption models defeats the original intention of saving energy in these works.

Since cellular networks with H2H/M2M coexistence are comprised of heterogeneous devices with different capabilities and characteristics as well as QoS requirements, the resource allocation processes are often hard to be well-defined mathematically, further reducing the effectiveness of traditional approaches to obtain optimal solutions. Therefore, intelligent resource management approaches have aroused interest of many researches due to their ability to provide specialized and customized resources for different IoT devices. In \cite{das2022reinforcement}, authors have designed an optimization problem that determines the optimal frequency and power allocation needed to maximize the achievable rate performance of all M2M and H2H users in the network subject to the co-tier interference and QoS constraints, and have proposed a RL-based resource utilization policy. Authors in \cite{xu2020deep} have introduced a DRL-based energy efficient resource allocation strategy for EH-M2M networks with consideration of the QoS of H2H traffic and the available energy in EH-devices. A joint power control and channel selection based on the DRL algorithm, combined with statistical channel state information (CSI), have investigated in \cite{zhao2019joint} to adaptively reduce interference. However, these intelligent methods have not clearly distinguish the M2M service types and ensure their QoS demands. In addition, the nonstationarity have not discussed while the developed RL scheme was a multi-agent algorithm.

Thus far, although a large number of researches of resource management have been carried out on cellular networks with H2H/M2M coexistence, they still have the following limitations: firstly, the potential of SWIPT to improve the EE of network has not been fully utilized. Secondly, the diverse QoS requirements of heterogeneous devices, and the interference caused by spectrum sharing have not been well considered. Finally, the resource allocation solutions using the traditional approaches were not globally optimal, while existing RL-based algorithms have not discussed the nonstationarity caused by multi-agent deployment.
\subsection{Contributions and Outline}
In this paper, we consider the energy-efficient resource allocation problem in context of the SWIPT enabled cellular network with H2H/M2M coexistence. The goal is to simultaneously achieve high EE performance and ensure divers QoS requirements by allocating the appropriate resources (i.e., spectrum, power and PS ratio) for tolerable M2M links. In system model, both critical machine-type and tolerable machine-type services integrated with SWIPT are considered in addition to H2H service. To support various QoS demands, we model the optimization problem as a multi-agent DRL formulation, where multiple tolerable M2M links attempt to share the frequency spectrum preoccupied by critical M2M and H2H links. To this end, we propose a multi-agent DRL-based autonomous spectrum-power-ratio allocation (MADRL-ASPRA) scheme by designing the states, actions and reward function, making agents accommodate the dynamic network environment and achieve the objective optimization. The main contributions of this work are as follows.
\begin{itemize}
  \item
  We put forward a sustainable network model of cellular networks with H2H/M2M coexistence, wherein the SWIPT enables MTCDs to harvest energy and decode information simultaneously. In addition to considering the existing conventional H2H users, newly introduced MTCDs are further subdivided into mission critical and mission tolerable type.

  \item
  We formulate the joint spectrum assignment, transmit power and PS ratio allocation problem with the goal of maximizing the EE of M2M links while considering the divers QoS constraints. For the QoS constraints, we explicitly model and separately solve the issues of meeting the H2H users' SINR thresholds, critical M2M links' outage ratios and tolerable M2M links' payload transmission probability.

  \item
  We model the EE optimization of multiple M2M links as a multi-agent problem and innovatively present a systematic approach for designing the state/action/reward functions to develop the MADRL-ASPRA scheme based on the nature of DRL. Specifically, the designed behaviour-tracking based state provides a stable training process and the common reward enables agents to work cooperatively in a distributed way. The mathematically exact expressions of QoS constraints are given via theoretical analysis, thereby reducing computational complexity.

  \item
  We verify the efficiency and superiority of proposed MADRL-ASPRA scheme by comprehensive simulations. Numerical results demonstrate that with the designed functions and proper training mechanism, our scheme outperforms other intelligent resource management approaches in terms of convergence speed and meeting the EE and diverse QoS requirements of heterogeneous users.
\end{itemize}

The rest of this paper is organized as follows: In the next section, we describe the system model and problem formulation. Section III present the proposed multi-agent DRL-based autonomous spectrum-power-ratio allocation algorithm. In section IV, we provide the evaluation results and finally section V will conclude the paper.
\section{SYSTEM MODEL AND PROBLEM FORMULATION}\label{System Model}
\subsection{Network Model}
In this work, we have considered a downlink SWIPT enabled H2H/M2M coexistence network consisting of a single BS, H2H user equipments (HUEs), machine type network gateways (MTCGs) and MTCDs, where the BS is deployed at the center of certain area, the HUEs and MTCGs are randomly deployed within the coverage area of the BS. Further, the MTCDs are distributed around the MTCGs and form several clusters. H2H links leverage cellular interfaces to connect all HUE to the BS for high data rate services, while the M2M links in one cluster correspond to local downlink communication between MTCDs and one MTCG.

Suppose that all MTCDs are equipped with the SWIPT technology for prolonging the operation lifetime of downlink signal transmission. By extracting energy of the received RF signals, the power can be effectively supplied to various MTCDs. All MTCDs are adopted with PS policy, which splits the received power into two parts for information decoding and EH. Similar to the works in \cite{muy2021energy}, each MTCD decodes its own information from the received signal, which contains all information for all MTCDs. Since the information transmitted to each MTCD has an identification information, each MTCD can decode its own information based on this identification information.
\begin{figure}[tb]
\begin{center}
\includegraphics[width=.4\textwidth]{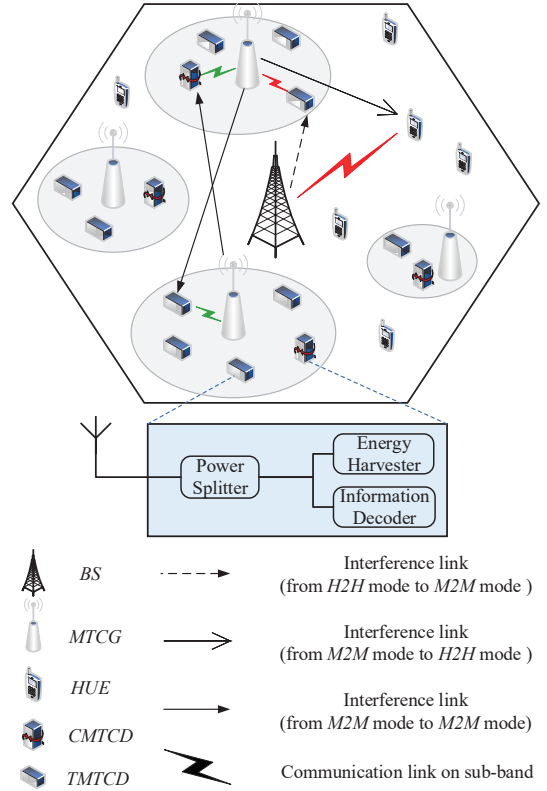}
\end{center}
\caption{System model of cellular network with H2H/M2M coexistence.}
\label{fig:1}
\end{figure}

The general orthogonal frequency division multiplexing (OFDM) is adopted to support multiuser transmission. Several consecutive subcarriers are grouped to form a spectrum sub-band \cite{liang2019spectrum}. We assume that each H2H link and critical M2M link have been preassigned an orthogonal spectrum sub-band with fixed transmit power. Since the same spectrum resource will be shared by multiple tolerable M2M links, the interference problems will arise. Accordingly, the research is to design an autonomous resource allocation scheme for tolerable M2M links such that both H2H and M2M links achieve their respective targets with maximal EE of M2M links.

Assume the set of available spectrum sub-bands within the network is indexed by $\mathcal{K}=\mathcal{H}\cup \mathcal{S}$, where $\mathcal{H}=\{1,2,\cdots,H\}$ represents HUE and $\mathcal{S}=\{1,2,\cdots,S\}$ represents critical machine type devices (CMTCDs). Moreover, let $\mathcal{M}=\{1,2,\cdots,M\}$ be the set of MTCGs or clusters, and the set of tolerable machine type communication devices (TMTCDs) is indexed by $\mathcal{N}=\{1,2,\cdots,N\}$. Finally, we introduce all SWIPT enabled MTCDs set as $\mathcal{D}=\mathcal{S}\cup \mathcal{N}$.
\subsection{Channel Model}
In this considered network, the channel between each link is composed of three parts, namely path loss, shadowing, and Rayleigh fading. In particular, path loss and shadowing are large-scale components of the channels, which remain unchanged for long time. Such the channel fading is supposed to be roughly the same across one sub-band. Assume that the communication link from the $m$-th MTCG to the $s$-th CMTCD and $n$-th TMTCD is denoted by $m,s$ and $m,n$, respectively. That is, during one coherence time period, the instantaneous channel gain between MTCG $m$ and TMTCD $n$ over sub-band $k$ is modeled as
\begin{equation}
g_{m, n}^{k}  = \chi_{m, n} \beta_{m, n} h_{m, n}^{k}
\end{equation}
where $\chi_{m, n}$ and $\beta_{m, n}$ are the frequency independent path loss and shadowing, respectively. $h_{m,n}^k$ is the frequency dependent small-scale fading power component. Similarly, we define $g_{m,s}^k$  as the the channel power gain from the MTCG $m$ the CMTCD $s$ over the sub-band $k$. $g_{BS,h}^k$ is the channel power gain from the BS to the HUE $h$ over the sub-band $k$. The interfering channel from the BS to the MTCD $n$ is denoted by $g_{BS,n}^k$. The interfering channel from the MTCG $m^{\prime}$ in other cluster to the TMTCD $n$ over the $k$-th sub-band is denoted by $g_{m',n}^k$, $m^{\prime}\in \mathcal{M}$, $m^{\prime}\neq m$. $g_{m',s}^k$, $m^{\prime}\in \mathcal{M}$, $m^{\prime}\neq m$ is the interfering channel from the MTCG $m^{\prime}$ in other cluster to the CMTCD $s$ over the $k$-th sub-band. $g_{m,h}^k$ represents the interfering channel from the MTCG $m$ to the HUE $h$ over the sub-band $k$. Table I is a summary of channel symbol notation.
\begin{table}[b]
\caption{Summary of Channel Symbol Notation}
\label{table:1}
\begin{center}
\begin{tabular}{p{30pt}p{180pt}}
\hline
\textbf{Symbol}  & \textbf{Definition}\\
\hline
$g_{m,n}^k$         & the channel from the $m$-th MTCG to the $n$-th TMTCD in same cluster over the $k$-th sub-band\\
\hline
$g_{m,s}^k$       & the channel from the $m$-th MTCG to the $s$-th TMTCD in same cluster over the $k$-th sub-band\\
\hline
$g_{BS,h}^k$      & the channel from the BS to the $h$-th HUE over the $k$-th sub-band\\
\hline
$g_{m^{\prime},n}^k$        & the interfering channel from the $m^{\prime}$-th MTCG to the $n$-th TMTCD in different cluster over the $k$-th sub-band\\
\hline
$g_{BS,n}^k$        & the interfering channel from the BS to the $n$-th TMTCD over the $k$-th sub-band\\
\hline
$g_{m,h}^k$ & the interfering channel from the $m$-th MTCG to the $h$-th HUE over the $k$-th sub-band\\
\hline
\end{tabular}
\end{center}
\end{table}
\subsection{Signal Model}
Each user may suffer from the co-channel interference including interference from one or multiple clusters if there exist other users using the same channel. To this end, the signal-to-interference-plus-noise ratios (SINRs) of the $h$-th H2H link and the $m,s$-th critical M2M link over the $k$-th sub-band can be respectively expressed as
\begin{equation}
\text{SINR}_{h}^{k}  = \frac{P_{B S, h} g_{B S, h}^{k}}{x_{m, n}^{k} \sum_{n \in \mathcal{N}} \sum_{m \in \mathcal{M}} P_{m, n}^{k} g_{m, h}^{k}+\sigma^{2}}
\end{equation}
and
\begin{equation}
\text{SINR}_{m, s}^{k}  =  \frac{\rho_{m, s} P_{m, s} g_{m, s}^{k}}{\rho_{m, s} I_{m, s}^{k}+\sigma^{2}}
\end{equation}
where $P_{B S, h}$ and $P_{m, s}$ on the molecule represent the emitting power of the BS and MTCG, respectively. $\sigma^2$ represents the additive white gaussian noise (AWGN) power. The first term in the denominator of equation (2) and (3) represent the interference from the MTCGs sharing the $k$-th sub-band with H2H or critical M2M links, $\rho_{m, s}\in[0,1]$ is the PS ratio at the $s$-th CMTCD. $x_{m,n}^k$ is the spectrum allocation indicator with $x_{m,n}^k=1$ implying the $m,n$-th tolerable M2M link is transmitting over the $k$-th sub-band and $x_{m,n}^k=0$ otherwise, The received interference $I_{m, s}^{k}$ of equation (3) includes the intra- and inter-cluster interference, and is represented as
\begin{equation}
I_{m, s}^{k}=x_{m^{\prime}, n}^{k} \sum_{n} \!\sum_{m^{\prime} \neq m} p_{m^{\prime}, n}^{k} g_{m^{\prime}, s}^{k}+x_{m, n}^{k} \sum_{n} p_{m, n}^{k} g_{m, s}^{k}
\end{equation}
where $p_{m, n}^{k}$ is the transmit power of the $m$-th MTCG to the $n$-th TMTCD over the $k$-th sub-band.

We evaluate the instantaneous SINR at each tolerable M2M link over the $k$-th sub-band as
\begin{equation}
\text{SINR}_{m, n}^{k}=\frac{\rho_{m, n}^{k} p_{m, n}^{k} g_{m, n}^{k}}{\rho_{m, n}^{k} I_{m, n}^{k}+\sigma^{2}}
\end{equation}
where $\rho_{m, n}^{k}$ is the PS ratio at the $n$-th TMTCD over the $k$-th sub-band. The received interference $I_{m, n}^{k}$ can be divided into two different spectrum sharing scenarios. When the spectrum sub-band k is occupied by the H2H link, $I_{m, n}^{k}$ can be expressed as
\begin{equation}
\begin{aligned}
I_{m, n}^{k}&=P_{m, s} g_{m, n}^{k}+x_{m^{\prime}, n}^{k} \sum_{n}\! \sum_{m^{\prime} \neq m} p_{m^{\prime}, n}^{k} g_{m^{\prime}, n}^{k}\\&+x_{m, n}^{k} \sum_{n} p_{m, n}^{k} g_{m, n}^{k}
\end{aligned}
\end{equation}

When the spectrum sub-band $k$ is occupied by the critical M2M link, $I_{m, n}^{k}$ is given by
\begin{equation}
\begin{aligned}
I_{m, n}^{k}&=P_{B S, n} g_{B S, n}^{k}+x_{m^{\prime}, n}^{k} \sum_{n} \!\sum_{m^{\prime} \neq m} p_{m^{\prime}, n}^{k} g_{m^{\prime}, n}^{k}\\&+x_{m, n}^{k} \sum_{n} P_{m, n}^{k} g_{m,n}^{k}
\end{aligned}
\end{equation}

Since the SWIPT technology can compensate the power consumption by harvesting surrounding energy, accordingly, the energy harvested via SWIPT at the $d$-th MTCD over the $k$-th sub-band can be described as
\begin{equation}
E H_{m, d}^{k}=\theta\left(1-\rho_{m, d}^{k}\right)\left(p_{m, d}^{k} g_{m, d}^{k}+I_{m, d}^{k}\right)
\end{equation}
where $d\in \mathcal{D}=\mathcal{S}\cup \mathcal{N}$, represents all the SWIPT enabled MTCDs, and $0<\theta \leq 1$ denotes the energy conversion efficiency.
\subsection{QoS Requirements}
Since the proposed network model contains HUE, tolerable and critical machine-type users with different capabilities and QoS requirements, it is natural to take these practical demands into account as constraints in a mathematical way. The traffic of H2H service is based on voice and internet communication, which puts forward higher requirements for data transmission rate. The most of M2M services are event-based, so their traffic is usually generated periodically with varying frequencies according to such M2M service's characteristics. As for critical M2M services, they tend to have stringent delay and transmission reliability requirements but are less interested in high data rates.

According to the above content, our scheme maintains the QoS requirements of conventional HUE by meeting their SINR thresholds ($\text{SINR}_{h}^{\min }$), i.e.,
\begin{equation}
\text{SINR}_{B S, h}^{k} \geq \text{SINR}_{h}^{\min }, \quad \forall k
\end{equation}

The QoS requirements of tolerable machine-type users are mathematically modeled as the packets of size $V$ delivered within a time budget $T$ as
\begin{equation}
\operatorname{Pr}\left\{\frac{\sum_{t=1}^{T} \sum_{k=1}^{K} x_{m, n}^{k} C_{m, n}^{k, t} \geq V}{\Delta_{\mathrm{T}}}\right\} \geq \overline{P_{n}}, \quad \forall k
\end{equation}
where $V$ refers to the size of the cyclicity generated M2M payload in bits, $\Delta_{\mathrm{T}}$ represents the channel coherence time. $C_{m, n}^{k, t}$ denotes the capacity of a particular M2M link $m,n$ over the sub-band $k$ can be calculated by Shannon formula as follows
\begin{equation}
C_{m, n}^{k, t}=B \log _{2}\left(1+\text{SINR}_{m, n}^{k, t}\right)
\end{equation}
where $B$ is the bandwidth of each spectrum sub-band. The purpose of adding index $t$ in $C_{m, n}^{k, t}$ is to express the capacity of such M2M link at different coherent time slots.

Generally, the QoS requirements of critical machine-type users can be represented as the outage probability, which can be captured by guaranteeing the received SINR at the CMTCD. The received SINR should not be above the target threshold ($\text{SINR}_{m,s}^{\min }$). Then, the reliability constraint is given by
\begin{equation}
\operatorname{Pr}\left\{\text{SINR}_{m, s}^{k} \leq \text{SINR}_{m, s}^{\min }\right\} \leq \overline{P_{s}}, \quad \forall k
\end{equation}
where $\overline{P_{s}}$ is the maximum tolerable outage probability.
\subsection{Problem Formulation}
The EE (bits/Hz/Joule) of all SWIPT enabled MTCDs is defined as the ratio of the total achievable spectrum efficiency to the total energy consumption \cite{zhou2019energy}, which is expressed as $\eta=\frac{R_{\text {total }}}{E C_{\text {total }}}$.

The total achievable spectrum efficiency $R_{\text {total }}$ is given by
\begin{equation}
R_{\text {total }}=\sum_{m} \sum_{d} \sum_{k} x_{m, d}^{k} \log _{2}\left(1+\text{SINR}_{m, d}^{k}\right), \quad \forall m, d, k
\end{equation}

The total energy consumption can be formulated as
\begin{equation}
\begin{aligned}
E C_{\text {total }} & = \sum_{m} \sum_{n} \sum_{k} p_{m, n}^{k} x_{m, n}^{k}+\sum_{m} \sum_{\mathrm{s}} P_{m, s} +P_{\mathrm{c}}\\
&-\sum_{m} \sum_{d} \sum_{k} x_{m, d}^{k} E H_{m, d}^{k}, \quad \forall m, n, k
\end{aligned}
\end{equation}
where $P_{C}$ is the circuit power consumption of all M2M links.

Therefore, the EE maximization problem should be formulated satisfying the diverse QoS requirements. Accordingly, the joint spectrum assignment, transmit power and PS ratio allocation problem in SWIPT enabled cellular networks with H2H/M2M coexistence is described as
\begin{align}
\max _{\rho\!_{m\!,\!n}^{k}\!,p\!_{m\!,\!n}^{k}\!,x\!_{m\!,\!n}^{k}}& \eta=\frac{R_{\text {total }}}{E C_{\text {total }}}\label{Formulation}\\
\operatorname{s.t.\quad} &\text{SINR}_{B S, h}^{k} \geq \text{SINR}_{h}^{\min }, \quad \forall h, k \tag{\ref{Formulation}{a}}\\
&\operatorname{Pr}\left\{\text{SINR}_{m, s}^{k} \leq \text{SINR}_{m, s}^{\min }\right\} \leq \overline{P_{s}}, \quad \forall s, k\tag{\ref{Formulation}{b}} \\
&\operatorname{Pr}\!\left\{\!\frac{\sum_{t} \sum_{k} x_{m, n}^{k} C_{m, n}^{k, t} \!\geq\! V}{\Delta_{\mathrm{T}}}\!\right\} \!\geq \!\overline{P_{n}},\quad \forall m, n\tag{\ref{Formulation}{c}}\\
&\rho_{m, n}^{k} \in[0,1], \quad \forall m, n, k \tag{\ref{Formulation}{d}}\\
& x_{m,n}^{k}\in\{0,1\}, \quad \forall m,n,k\tag{\ref{Formulation}{e}}\\
& p_{m,n}^{k}\leq \overline{P}, \quad \forall m,n,k\tag{\ref{Formulation}{f}} \\
& \sum_{k} x_{m,n}^{k}\leq 1, \quad \forall m,n \tag{\ref{Formulation}{g}}
\end{align}
where $\rho_{m,n}^{k}$, $p_{m,n}^k$ and $x_{m,n}^{k}$ represent the PS ratio, transmit power allocation and spectrum sub-band assignment strategy, respectively. Constraints (15a), (15b) and (15c) contain respectively the QoS requirements of H2H, critical M2M and tolerable M2M links. Constraint (15d) divides $\rho_{m,n}^{k}$ $\left(0\leq \rho_{m,n}^{k} \leq 1\right)$ portion of the signal power to information decoding while the remaining $1-\rho_{m,n}^{k}$ portion of power to EH. Constraint (15e) defines the binary variable to indicate the mapping between sub-band and user. Constraint (15f) ensures the transmit power of tolerable M2M links lower than their maximum limits $\overline{P}$. Constraint (15g) models our assumption that each of the tolerable M2M link accesses a single spectrum sub-band.

Since the optimization goal advocates for a harmonious coexistence between diverse devices, the nonlinear constraints of diverse devices make the objective function a NP-hard combination problem. Therefore, the spectrum assignment, transmit power and PS ratio allocation solutions obtained by using the traditional resource management approaches are not globally optimal.

To this end, considering the fact that multiple tolerable M2M links attempt to access limited spectrum occupied by H2H and critical M2M links, we model the EE optimization of multiple M2M links as a multi-agent problem, then develop a DRL-based autonomous resource allocation approach to learn the optimization policy under divers QoS constraints in the next section.
\section{MULTI-AGENT DRL BASED RESOURCE ALLOCATION}\label{Section_III}
\subsection{Related Definition of Multi-Agent DRL}
DRL provides an autonomous decision-making mechanism for the network entities by combining both RL and DNN, where the RL focuses on how agents ought to take actions in the process of interacting with environment such that some notion of cumulative reward is maximized. The environment is typically formulated as a Markov decision process (MDP). MDP is defined by a tuple $\left(S_n,A_n,P_{s,s^{\prime}},R_n\right)$, where $S_n$ denotes a finite set of environment states of the $n$-th agent, $A_n$ denotes a finite set of agent actions and $P_{s,s^{\prime}}$ is the probabilistic transition from one state to another. $R_n$ is the reward function, which reflects the learning goal and is fed back to the agent by the environment. The ultimate goal is to find the optimized policy $\pi^{\ast}:S \rightarrow A$ that maximizes the cumulative discounted reward at time $t$. The mathematic expression is
\begin{equation}
G_t = \sum_{n=0}^{\infty} \gamma^{n}R_{t+n+1}
\end{equation}
where $\gamma$ is the discount factor satisfying $0\leq \gamma \leq 1$.

Q-learning \cite{watkins1992q} is among the most popular RL methods to address the MDP problems. The core behind this classical method is to define the infinite execution of learning procedure as a mathematical expectation. The specific description is to assume a stationary setting, where the Q-function represents the value of the policy $\pi$ for selecting action $a$ at state $s$, i.e.,
\begin{equation}
Q^{\pi}(s, a)=\mathbb{E}_{\pi}\left\{G_{t} \mid S_{t}=s, A_{t}=a\right\}
\end{equation}

As described in \cite{saraiva2020deep}, with a variant of the stochastic approximation condition on the learning rate and the assumption that all state-action pairs continue to be updated, the learned action-value function in Q-learning converges with probability 1 to the optimal $Q^{\pi^{\ast}}$. Hence, the Q-value can be updated recursively as follows
\begin{equation}
Q(s, a) \leftarrowtail Q(s, a)+\alpha\left[R+\gamma \max _{a \in A} Q\left(s^{\prime}, a\right)-Q(s, a)\right]
\end{equation}
where $\alpha$ is the learning rate. However, the classical Q-learning uses a look-up table to represent Q-values and the computational complexity increases along with the size of the states and action spaces. The simple look-up table where separate Q-value is maintained for each state/action pair is not feasible in large space with massive number of states.

To combat this problem, DNN is adopted to replace the look-up table in the classical Q-learning algorithm, which is essentially a DRL-based algorithm to estimate the Q-value function. Therefore, a DNN function approximator called DQN can be expressed as $Q(s,a;\omega)$ with $\omega$ being its weight. The input layer of DQN is the environment states, and each port of the output layer provides the approximated Q-value function $Q(s,a;\omega)$. The core concept of DQN is that the function $Q(s,a;\omega)$ is completely determined by $\omega$. Consequently, the task of finding the best Q-function is essentially limited to the search for these best parameters of finite dimensions \cite{foerster2017stabilising}.

At each training step, in a similar vein, the agents retain their experiences of interacting with environment and store them in the experience replay memory $D$, which collects and stores the experiences until time $t$ in the form of transition tuple $(S_t,A_t,R_{t+1},S_{t+1})$. The use of experience replay can stabilize learning and prevent oscillations or divergence. Moreover, DQN builds an additional target $Q$ network $Q^{-} (s,a;\omega^{-})$ as the "quasi-static target network" method implies \cite{mnih2015human}, which alongside the training network to stabilize the overall network performance. With a random sample minibatch from the experience replay memory and target $Q$ network, each DQN agent can update its weights to minimize the loss function given by
\begin{equation}
L(\omega)=\mathbb{E}_{\left(S_{t}, A_{t}, R_{t+1}, S_{t+1}\right) \in \mathcal{D}}\left\{\left[y_{t}^{D Q N}-Q\left(S_{t}, A_{t} ; \omega\right)\right]^{2}\right\}
\end{equation}
where $y_{t}^{DQN} $is the target value estimated by
\begin{equation}
y_{t}^{DQN} = R_{t+1} + \gamma \max_{a^{\prime}} Q^{-}\left( S_{t+1},a^{\prime},\omega^{-} \right)
\end{equation}

The successful transformation of deep learning on single-agent RL into multi-agent setup with the help of one classical Q-learning algorithm, namely independent Q learning (IQL). As the most popular multi-agent RL method, IQL makes each agent independently learn its own policy, and treat other agents as part of the environment \cite{tan1993multi}. But from the perspective of each agent, the environment becomes unstable as other agents learn and adjust their behaviour at the same time. Fortunately, substantial empirical evidence has shown that IQL often performs well in practice. As mentioned earlier, DQN especially relies on the experience reply to improve stability. However, the combination of the experience replay with IQL seems to be problematic: the generated experiences in replay memory no longer reflects the dynamic changes of the current environment, and the returned experiences are mostly outdated, thus confusing the agents. Consequently, how to avoid nonstationarity in multi-agent scenarios is now a matter of active research. To solve this problem, we will utilize the definitions mentioned above and describe the design of the proposed algorithm in detail in next subsection.
\subsection{Proposed Multi-Agent DRL-based Autonomous Spectrum-Power-Ratio allocation Algorithm}
In our resource management scenario, each tolerable M2M link operates as a learning agent and maintains its own DQN with weights $\omega$ and $\omega^{-}$. Each agent obtains experiences through interaction with the unknown environment, which are then used to guide its own policy design. Multiple M2M agents explore the environment at the same time, and refine spectrum, power and PS ratio allocation strategies based on their environment states. Although the resource management problem may seem like a competitive game, we convert it to a fully cooperative one by using the same reward for all agents on behalf of global network performance.

The developed multi-agent DRL-based autonomous spectrum-power-ratio allocation (MADRL-ASPRA) algorithm is implemented in two stages, namely training stage and testing stage. In the training stage, each agent can easily obtain the global network performance-oriented reward, and then realigns its actions to approach the optimal policy by updating its own NN weights. In the testing stage, each agent observes the network environment states and then selects the actions according to its trained DQN on a time scale on par with the small-scale channel fading. In order to give a minute description of our proposed scheme, we need to define the state, action and reward function.

$\boldsymbol{State}$: As the basic for policy-making, states should provide enough information for agents to learn optimal policy. Therefore, in each time slot, the current environment state includes global channel information and all agents' behaviours. For clearness, we define the state space for $m,n$-th tolerable M2M link (agent) as follows
\begin{equation}
\begin{aligned}
S_{m, n}^{t}=\bigg\{&\big\{G_{m, n}^{k}\big\}_{k \in \mathcal{K}},\big\{I_{m, n}^{k}\big\}_{k \in \mathcal{K}}, V_{m, n}, T_{m, n},\\
&\big\{\text{QoS}_{h}\big\}_{h \in \mathcal{H}},\big\{\text{QoS}_{s}\big\}_{s \in \mathcal{S}}\bigg\}
\end{aligned}
\end{equation}
which contains both local and nonlocal information. $\left\{G_{m,n}^{k}\right\}_{k\in \mathcal{K}}$ is the channel information observed by agent $m,n$ and consists of the following sets: target channel $g_{m,n}^{k}$ for all $k\in \mathcal{K}$, interfering channel $g_{BS,n}^{k}$ and $g_{m^{\prime},n}^{k}$ for all $k\in \mathcal{K}$ and $m^{\prime}\in \mathcal{M}, m^{\prime}\neq m$, the channel from its own transmitter to the $s$-th CMTCD $g_{m,s}^{k}$ for all $k\in \mathcal{K}$ and $s\in \mathcal{S}$. $\left\{I_{m,n}^{k} \right\}_{k\in \mathcal{K}}$ is the received interference over all sub-bands. $V_{m,n}$ and $T_{m,n}$ denote the remaining transmission payload and time budget of $m,n$-th agent, respectively. $\left\{\text{QoS}_{h} \right\}_{h\in \mathcal{H}}$ and $\left\{\text{QoS}_{s} \right\}_{s\in \mathcal{S}}$ are the QoS satisfaction indicators, indicating the satisfied degree of QoS requirements. $\left\{\text{QoS}_{h} \right\}_{h\in \mathcal{H}} = 1$ indicating the QoS requirements of all H2H links are guaranteed and $\left\{\text{QoS}_{h} \right\}_{h\in \mathcal{H}} = 0$ otherwise. Similarly, $\left\{\text{QoS}_{s} \right\}_{s\in \mathcal{S}}=1$ utilizes same indicators to indicate the QoS satisfaction levels of all critical M2M links.

Furthermore, to address the nonstationarity problem mentioned above, we draw lessons from \cite{foerster2017stabilising}. Since the value function of each agent is mapped by the high dimensional DQN weights, it is undesirable to take all the DQN parameters of other agents as the input to the value function. One of the methods in \cite{foerster2017stabilising} is to condition each agent's value function on a fingerprint that disambiguates the age of the experiences sampled from the replay memory. The kernel of this method is to determine a low-dimensional fingerprint that correlated with the true value of state-action pairs. In our research, the training iteration number $e$ and the rate of exploration $\epsilon$ can be the fingerprint to track the behaviour of the agents. This approach is feasible because the nonstationarity of each agent's environmental dynamics is affected by changes in the polices of other agents, instead of by the policies themselves. Eventually, we can mix in fingerprint to augment the state space as follows
\begin{equation}
\begin{aligned}
S_{m, n}^{t}=\bigg\{&\big\{G_{m, n}^{k}\big\}_{k \in \mathcal{K}},\big\{I_{m, n}^{k}\big\}_{k \in \mathcal{K}}, V_{m, n}, T_{m, n},\\
&\big\{\text{QoS}_{h}\big\}_{h \in \mathcal{H}},\big\{\text{QoS}_{s}\big\}_{s \in \mathcal{S}},e,\epsilon\bigg\}
\end{aligned}
\end{equation}

$\boldsymbol{Action}$: The resource management of SWIPT enabled cellular network with H2H/M2M coexistence comes down to the spectrum sub-band assignment, transmit power and PS ratio allocation for each tolerable M2M link. The $m,n$-th agent selects action set $A_{m,n}^{t}$ under the current observed state $S_{m,n}^{t}$. The action set $A_{m,n}^{t}$ is defined as
\begin{equation}
A_{m,n}^{t} = \left\{o_{m,n},p_{m,n}^{k},\rho_{m,n}^{k} \right\}
\end{equation}
where $o_{m,n}\in \left\{1,2,\cdots,K\right\}$ denotes the selected spectrum sub-band, transmit power selection $p_{m,n}^{k}\in \left\{\frac{\overline{P}}{L},\frac{2\overline{P}}{L},\cdots,\overline{P}  \right\}$ and PS ratio selection $\rho_{m,n}^{k}\in \left\{\frac{1}{Z},\frac{2}{Z},\cdots,1 \right\}$ are defined as $L$ and $Z$ discrete levels, respectively.

At each time slot, each agent observes the environment state and selects the actions with a particular decision-making strategy. The $\epsilon$-greedy strategy is a simple and effective one, which is described as follows
\begin{equation}
a^{\prime}\!=\!\begin{cases}
\text { random action, } & \text { with probability } \epsilon \\
\underset{a^{\prime}}{\operatorname{argmax}} Q_{m, n}\!\left(S_{m, n}^{t}, a^{\prime} ; \omega\right),& \text { with probability } 1-\epsilon
\end{cases}
\end{equation}
where each agent explores or exploits the environment by taking random actions or greedy actions according to a given $\epsilon$ probability distribution, respectively \cite{saraiva2020deep}.

$\boldsymbol{Reward}$: The design of reward function is the most charming and flexible link in DQN, and also the key to endow the algorithm with intelligence to handle difficult optimization problems. Since the reward function directly affects the objectives of the agents, the design should be related to the desired objective. In the EE optimization problem formulated in section II, we hope to achieve the following goals by designing the reward function: maximizing the overall EE of SWIPT enabled MTCDs while ensuring the high transmission rate of HUE, high reliable communication of critical M2M links and improving the payload transmission success rate of tolerable M2M links.

To this end, first we simply include the instantaneous sum EE of all M2M links $\eta^{t}$ as defined in (15) in the reward function at each time slot $t$. Then the QoS assurance expressions take second place, i.e., keep HUE's SINR above threshold $\text{SINR}_{h}^{min}$ as defined in (9), maintain TMTCDs' payload transfer-rate above lower probability $\overline{P_{n}}$ as defined in (10), and maintain CMTCDs' outage ratio below tolerable outage probability $\overline{P_{s}}$ as defined in (12).

Obviously, the SINR of HUE in each time slot can be easily obtained from the received power and interference. However, the QoS constraints of MTCDs are mathematically modeled as probabilities. Due to the absence of analytical expressions, agents need to randomly generate a large number of analog channels to obtain the probabilities in each time slot. Consequently, a variety of computation resources are being consumed and the learning process is restricted seriously. In response to this problem, we need to turn these QoS constraints into analytical forms.

Firstly, we set the reward $U_{n}$ equal to M2M transmission time remaining until the payload is delivered, after which the reward is set to a constant number. As such, the tolerable M2M-related reward at time slot $t$ is given by
\begin{equation}
U_{k}^{t}=\begin{cases}
\frac{T-t}{T}, \quad &\sum_{k=1}^{K}x_{m,n}^{k}C_{m,n}^{k,t}<V \\
0, \quad&\text { otherwise }
\end{cases}
\end{equation}

Through this transformation, agents can take into account both the transmission time remaining and payload delivery rate in the learning process.

Secondly, we use lemma and mathematical manipulations to calculate the outage probability of CMTCDs. When the $s$-th CMTCD shares the $k$-th sub-band with tolerable M2M links in different clusters, equation (13) can be replaced by
\begin{equation}
\operatorname{Pr}\left\{\frac{\rho_{m, s} P_{m, s} g_{m, s}^{k}}{\rho_{m, s} I_{m, s}^{k}+\sigma^{2}} \leq \text{SINR} _{m, s}^{\min }\right\} \leq \overline{P_{s}}
\end{equation}

Drawing on the following results in \cite{kandukuri2002optimal}, the analytical form of outage constraint (26) can be obtained.
\newtheorem{lemma}{Lemma}
\begin{lemma}
\textit{Suppose $\left\{z_1,\cdots,z_n\right\}$ are independent exponentially distributed random variables with means $\mathbb{E}[z_i]=1/{\lambda_i}$. Then, we have}
\begin{equation}
\operatorname{Pr}\left\{z_{1} \leq \sum_{i=2}^{n} z_{i}+c\right\}=1-e^{-\lambda_{1} c} \prod_{i=2}^{n}\left(\frac{1}{1+\frac{\lambda_{1}}{\lambda_{i}}}\right)
\end{equation}
\textit{where $c$ is a positive constant.}
\end{lemma}

With the aforesaid lemma, the outage constraint for each CMTCD in (26) can be replaced with
\begin{equation}
\begin{aligned}
1-\bigg[&\exp \bigg(\frac{\sigma^{2} \text{SINR}_{m, s}^{\min }}{\rho_{m, s} P_{m, s} \chi_{m, s} \beta_{m, s}}\bigg)\prod_{\dot{m}} \prod_{n}\\ &\bigg(1+\frac{\dot{I_{m, s}^{k}} \text{SINR}_{m, s}^{\min }}{P_{m, s} \chi_{m, s} \beta_{m, s}}\bigg)\bigg]^{-1}\leq \overline{P_{s}}
\end{aligned}
\end{equation}

$\dot{m}$ in the above formula can be either $m$ or $m^{\prime}$ according to the suffered interference, which is used to indicate the inter- or intra-cluster interference. $\dot{m}$ can also contains both $m$ and $m^{\prime}$, which means the interference includes both inter- and intra-cluster interference. Worthy of note is that only $\prod_{n}$ remained in the above inequality when the interfering signal is intra-cluster interference. Correspondingly, $\dot{I_{m,s}^{k}}$ denotes the received interfering signal power of $s$-th CMTCD over the $k$-th sub-band, which is given by
\begin{equation}
\dot{I_{m, s}^{k}}=\begin{cases}
p_{m,n}^{k} \chi_{m,s} \beta_{m,s}, &\dot{m}=m \\
p_{m^{\prime},n}^{k} \chi_{m^{\prime},s} \beta_{m^{\prime},s}, &\dot{m}=m^{\prime}\\
\end{cases}
\end{equation}

Besides, when $\dot{m}$ contains both $m$ and $m^{\prime}$, the replaced outage constraint by (28) needs to be obtained by multiplying every cases in (29) simultaneously.

Obviously, the replaced outage constraint is still a complicated formula, particularly when the interference includes both inter- and intra-cluster interference. To reduce the high computational complexity caused by multiplication calculation, we further reference to the following conclusions in \cite{papandriopoulos2005optimal} to bound the derived outage constraints
\begin{equation}
\begin{aligned}
&1-\bigg[\exp \bigg(\frac{\sigma^{2} \text{SINR}_{m,s}^{\operatorname {min}}}{\rho_{m, s} P_{m, s} \chi_{m, s} \beta_{m, s}}\bigg) \prod_{\dot{m}} \prod_{n} \\&\qquad \bigg(1+\frac{\rho_{m, s}^{k} \dot{I_{m, s}^{k}} \text{SINR}_{m, s}^{\min }}{\rho_{m, s} P_{m, s} \chi_{m, s} \beta_{m, s}}\bigg)\bigg]^{-1} \\
&\leq 1-\exp \bigg[-\frac{\text{SINR}_{m, s}^{\min }\bigg(\ddot{I_{m, s}^{k}}+\sigma^{2}\bigg)}{\rho_{m, s} P_{m, s} \chi_{m, s} \beta_{m, s}}\bigg] \\
&\leq \overline{P_{s}}
\end{aligned}
\end{equation}
where$\ddot{I_{m, s}^{k}}=x_{m^{\prime}, n}^{k} \sum_{n} \sum_{m^{\prime} \neq m} p_{m^{\prime}\!, n}^{k} \chi_{m^{\prime}\!, s}^{k}\beta_{m^{\prime}\!, s}^{k}+x_{m, n}^{k} \sum_{n} \\p_{m, n}^{k} \chi_{m, s}^{k}\beta_{m, s}^{k}$. Tightness of the upper bound on outage probability has been demonstrated in \cite{papandriopoulos2005optimal}. Therefore, the outage probability can be obtained by using equation (30) with lest computation cost. To enable agents to clearly distinguish how good the achieved QoS is, we further set reward $U_{s}^{t}$ to capture the QoS satisfaction of CMTCD $s$ at time slot $t$ which is given by
\begin{equation}
U_{s}^{t}=\begin{cases}
0, \quad p_{s}\leq \overline{P_{s}} \\
1, \quad\text { otherwise }
\end{cases}
\end{equation}
where $p_{s}$ is outage probability of $s$-th CMTCD calculated by equation (30). As the same, the indicator of QoS satisfaction $U_{h}^{t}$ of $h$-th HUE at time slot $t$ is expressed as
\begin{equation}
U_{h}^{t}=\begin{cases}
0, \quad \text{SINR}_{BS,h}^{k}\geq \text{SINR}_{h}^{\min} \\
1, \quad\text { otherwise }
\end{cases}
\end{equation}

\begin{algorithm}[t]
  \caption{The Multi-Agent DRL-Based Autonomous Spectrum-Power-Ratio Allocation Algorithm}
  \begin{algorithmic}[1]
   \State $Input:$ DQN structure, environment simulator, episode times $N_{e}$, synchronization times $N_{\omega}$
   \For{each M2M agent}
     \State Initialize DQN weights $\omega$ and $\omega^{-}$ randomly
     \State Initialize replay memory $D$
   \EndFor
   \For {$e$ to $N_e$}
      \State Reset payload $V$ and time budget $T$ for all agents, let $t$=1
    \Repeat
	\For{each M2M agent}  	
      	  \State Receive initial observation state $S^{t}_{m,n}$	
	  \State Select the action $A^{t}_{m,n}$  at the state $S^{t}_{m,n}$ by using $\epsilon$-greedy policy 	
      	\EndFor	
     \State Let $t=t+1$
	 \State All agents perform selected actions and receive an immediate reward $R^{t+1}$
	 \State Update channel small-scale fading
	 \For{each M2M agent}  	
      	  \State Observe the new state $S^{t+1}_{m,n}$	
	  \State Store transition  $(S^{t}_{m,n},A^{t}_{m,n},R^{t+1},S^{t+1}_{m,n})$ in $D_{m,n}$
	 \EndFor	
    \Until $t>T$
    \For{each M2M agent}  	
      	  \State Randomly sample minibatches from $D_{m,n}$
	  \State Update the DQN weights by using stochastic gradient descent to minimize the loss as defined in  (19)
    \EndFor
     \If {$e$ mod $N_{\omega}$ == 0}
      \State Let $\omega^{-}$= $\omega$
      \EndIf
        \EndFor
\State $output:$ Learned DQN
  \end{algorithmic}
\end{algorithm}

As a result, we define the reward at each time slot $t$ as
\begin{equation}
R^{t+1}=\mu \eta^{t} - U_{n}^{t}-U_{s}^{t} - U_{h}^{t}
\end{equation}
where $\mu$ is a positive weight to balance the gain and cost. As a result, the common reward function allows agents to work cooperatively to improve the EE of all M2M links while taking into account such constraints as SINR threshold, outage ratio and payload transmission probability.

During the resource allocation process, the spectrum naturally breaks into $K$ disjoint sub-bands, which is occupied by H2H link or critical M2M link accordingly. In each iteration, we trade on random sample minibatches from replay memory to renew the multiple DQN weights, until achieving the purpose of effective learning resource allocation policy. The pseudo code of the proposed multi-agent DRL-based autonomous spectrum-power-ratio allocation algorithm is summarized in Algorithm 1.

The proposed scheme will first be deployed to interact with an unknown H2H/M2M coexistence environment and learn optimal joint spectrum-power-ratio allocation policy from the interaction. After the expensive training, the well trained DQN will be deployed in the actual implementation to guide agent make reasonable resource allocation decisions without additional training and operations. Moreover, the trained DQN can still maintain high EE performance in the inexpensive testing stage without changing environment significantly.

\section{PERFORMANCE EVALUATION}\label{performance evaluation}
In this section, we present simulation results to verify the proposed multi-agent DRL based autonomous spectrum-power-ratio allocation algorithm (MADRL-ASPRA) for SWIPT enabled H2H/M2M coexistence network. We compare the proposed resource allocation scheme with the following schemes:
\begin{enumerate}
  \item
  The classical multi-agent Q-learning based scheme in \cite{das2022reinforcement}, termed MAQL, in which each state/action pair has a separate Q-value. In each iteration, each agent finds a policy maximizing its Q-value in the table look-up way.

  \item
  The single-agent DRL based scheme in \cite{xu2020deep}, termed SADRL, where only one agent updates its action selections in each iteration according to the instantaneous observation and trained DQN. One single DQN is used by all agents.

  \item
  The non-SWIPT-MADRL, where the M2M receivers do not have the EH function. Therefore, in order to verify the improvement of system performance brought by SWIPT, we further benchmark the proposed algorithm in the network environment without SWIPT enabling.
\end{enumerate}

These resource management schemes can be broadly divided into two categories: DRL-based schemes and RL-based scheme. The DRL-based schemes include MADRL-ASPRA, SADRL and non-SWIPT-MADRL. The RL-based scheme is MAQL.
\begin{table}[t]
\caption{Parameters of Simulation}
\label{table:2}
\begin{center}
\begin{tabular}{|m{120pt}<{\centering}|m{100pt}<{\centering}|}
\hline
\textbf{Parameter}  & \textbf{Value}\\
\hline
TMTCDs per cluster & Variable \\
\hline
Bandwidth of system&4MHz\\
\hline
Path loss &-128-37.6$\log_{10}{d}$, $d$ in km\\
\hline
Shadowing& Log-normal distribution with standard deviation of 8 dB\\
\hline
Fast fading& Rayleigh fading\\
\hline
Noise power $\sigma^{2}$&-114 dBm\\
\hline
Circuit power $p_{c}$&10 dBm\\
\hline
Energy conversion factor $\theta$&0.7\\
\hline
H2H transmit power $P_{BS,h}$ &30 dBm\\
\hline
Critical M2M transmit power $P_{m,s}$ &23 dBm\\
\hline
Tolerable M2M maximum transmit power $\overline{P}$ &Variable\\
\hline
SINR threshold for H2H $\text{SINR}^{\min}_{h}$&7 dB \\
\hline
SINR threshold for critical M2M $\text{SINR}^{\min}_{s}$&5 dB \\
\hline
Reliability for critical M2M $\overline{P}_{s}$&0.01 \\
\hline
Time constraint of tolerable M2M payload transmission $T$&100 ms\\
\hline
Payload size $V$ &3$\times$1024 Bytes\\
\hline
\end{tabular}
\end{center}
\end{table}

\subsection{Simulation Setup}
We consider an isolated cell with a radius of 500 (m), with the BS at the center, 2 HUEs and 2 MTCGs uniformly and randomly distributed in the cell. Each MTCG is surrounded by several MTCDs which are no more than 30 (m) away from the MTCG. And Each cluster composed of M2M devices contains one critical M2M link. We discretize transmission power and PS ratio into 10 and 5 levels, i.e., $L$=10, $Z$=5, respectively. Besides, we consider the practical effects of finite-precision digital processing, which limits the maximum value of the received SINR to 30dB. Other simulation parameters are given in Table II. By default, the TMTCDs per cluster and tolerable M2M maximum transmit power $\overline{P}$ are 2 and 15 dBm, respectively, and the settings in each figure take precedence wherever applicable.

With respect to the proposed scheme, the DQN model consists of three fully-connected hidden layers, and each contains $K\times L\times Z$ neurons which is the number of optional actions. The activation function is the rectified linear unit (ReLU): $f(x)=\max(0,x)$. RMSProp optimizer is adopted to minimize the loss function as defined in (19) with a learning rate of 0.001. The $\epsilon$-greedy strategy is used for first 80\% of iterations with $\epsilon$ linearly annealed from 1 to 0.01, and then $\epsilon$ remain unchanged in last 20\%. The target DQN is synchronized with the trained DQN every 4 iterations, i.e., $N_{\omega}$ = 4. The positive weight $\mu$ = 1/50.
\subsection{Convergence Comparisons}
We first evaluate the effects of the number of iterations on the performance of the proposed scheme. In particular, considering the features offered by each algorithm, the executive processes are discussed in two parts as follows. In Fig.2(a), it can be seen that the DRL-based schemes learn from scratch and have adequately trained for 8000 iterations. And after training, the DRL-based schemes are transferred to testing along with the RL-based scheme, which results in Fig.2(b) after 8000 iterations.

Accordingly, Fig.2(a) shows that for a fixed network environment simulator, the proposed MADRL-ASPRA, SADRL and non-SWIPT-MADRL schemes start converging after about 600, 6000 and 200 training iterations, respectively. This is attributed to the fact that optimal weights of DQN and policy have been obtained in those iterations. Besides, in contrast to the single-agent scheme, the multi-agent schemes not only convergence faster, but also has less fluctuations, which results from the intrinsic rule of the single-agent scheme. For the single-agent scheme, only one agent updates its action in each time slot while other agents' actions remain unchanged. In a nutshell, SADRL scheme requires more iterations of training to find a policy that fits all agents, as most of agents' action selections are not optimal in some time slots. Moreover, non-SWIPT-MADRL scheme has the fastest speed and most stable convergence process due to its action space is smaller without PS ratio allocation.
\begin{figure}[!t]
\begin{center}
\subfigure[training]{
\includegraphics[width=.4\textwidth]{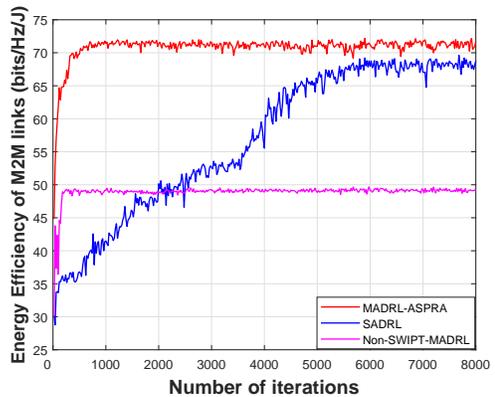}}
\subfigure[testing]{
\includegraphics[width=.4\textwidth]{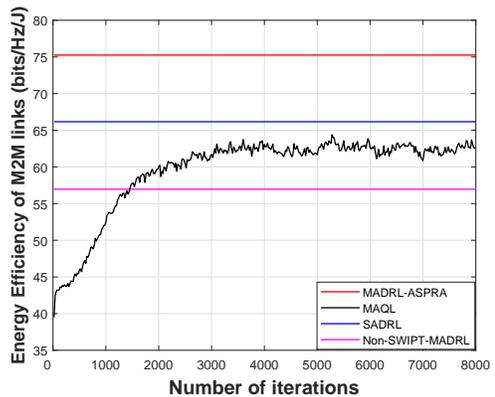}}
\caption{Aggregate EE achieved for each episode with increasing iterations.}
\end{center}
\label{fig:2}
\end{figure}

From Fig.2(b), it can be seen that the EE performance remain well without further training for DRL-based schemes. However, the RL-based scheme requires up to 4000 iterations to find the optimal policy. The fluctuations of performance of the RL-based scheme are higher than that of others due to the dynamic change of the observations of environment state. The results demonstrate that the DRL-based schemes have distinct advantages over RL-based algorithm. That is, the well trained DQN can be applied to address the problems on an ad hoc basis, whereas the QL needs an expensive iterative process at each problem solving session. Moreover, the SWIPT enabled schemes have obvious superiority in improving EE compared to the non-SWIPT enabled scheme, in which our approach achieves the highest EE performance.
\subsection{Energy Efficiency Performance Comparisons}
Fig.3 illustrates the aggregate EE of M2M links against the number of MTCDs in the network. It is noticeable that as the MTCD grows, the EE performance first increases and then tends to decline. We demonstrate the following aspects in order to better explain the simulation results.
\begin{figure}[t]
\begin{center}
\includegraphics[width=.4\textwidth]{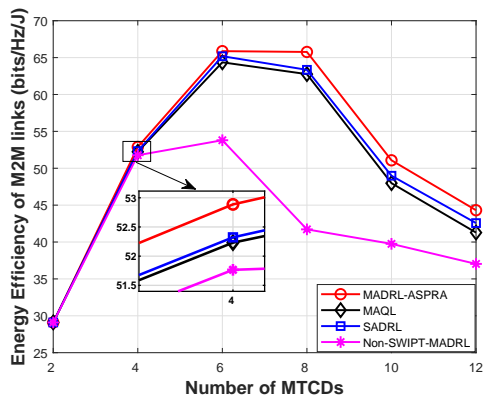}
\end{center}
\caption{Aggregate EE performance of M2M links with varying MTCDs. TMTCDs per cluster = 1, 2, 3, 4, 5.}
\label{fig:3}
\end{figure}

\noindent -- The number of MTCDs is 2: there is little difference in aggregate EE for either SWIPT-enabled or non-SWIPT-enabled scheme.  This is because when only CMTCDs and HUEs exist, the sub-bands they preoccupy are orthogonal without interference within the district. And owing to the geographical superiority, the channel conditions between each critical M2M links are good, further contributing to high SINR. Moreover, since CMTCDs use most of received power for information decoding, the amount of EH is too small to be measured.

\noindent -- The number of MTCDs from 2 to 6: the aggregate EE of M2M links of all schemes is improved accordingly, eventually reaching a maximum at 6 MTCDs. The reason is that spectrum reuse can improve the spectrum efficiency so that the EE is improved further. When the number of MTCDs is 6, 4 TMTCDs are introduced on the basis of 2 CMTCDs. That is, 4 tolerable M2M links attempt to access limited spectrum preoccupied by H2H and critical M2M links, so that each scheme can allocate one occupied spectrum sub-band to each tolerable M2M link without generating intra cluster interference which will seriously affect network performance. As a result, the highest EE performance is facilitated when the spectrum efficiency of the network reaches a maximum.

\noindent -- The number of MTCDs from 6 to 12: the aggregate EE of M2M links of all schemes is progressively diminishing as the MTCD density increases. This occurs due to excessive spectrum reuse incuring both inter- and intra-cluster interference. In this case, the interference becomes more pervasive and the power consumption increases dramatically, ultimately resulting in a degradation in EE performance.

\noindent -- Algorithm comparisons based on enabling conditions: the EE performance of non-SWIPT scheme deteriorates remarkably when the number of MTCDs exceeds 6, while the other SWIPT schemes can maintain higher EE performance until the number of MTCDs exceeds 8. This is because compared with non-SWIPT scheme, SWIPT schemes can convert strong interference signals into EH quantities despite the slight decrease in spectral efficiency. Therefore, on the basis of tradeoff between spectrum efficiency and EH, SWIPT schemes can achieve fairly good performance under certain interference intensity. And Such performance degradation becomes less dramatic as the number of MTCDs continues to increase. This is due to the fact that the M2M link suffers from severe interference, resulting in a significant impact on spectral efficiency, which leaves very little space for further performance degradation.

\noindent -- Algorithm comparisons: our proposed scheme not only can achieve the highest EE performance, but also has a better property of tradeoff, which is reflected in the fact that there is almost no decline in the aggregate EE when the number of MTCDs increases from 6 to 8. The reason is that the multi-agent setup decentralizes the training process, so that each agent can learn the optimal policy more efficiently. However, the decision-making in the centralized training process of SADRL needs to take into account the performance of all agents at each time slot, which accordingly causes the performance loss. Nevertheless, the MAQL scheme is inferior to the DRL-based schemes due to its cumbersome table look-up approach, which cannot take all strategies into account.

\begin{figure}[t]
\begin{center}
\includegraphics[width=.4\textwidth]{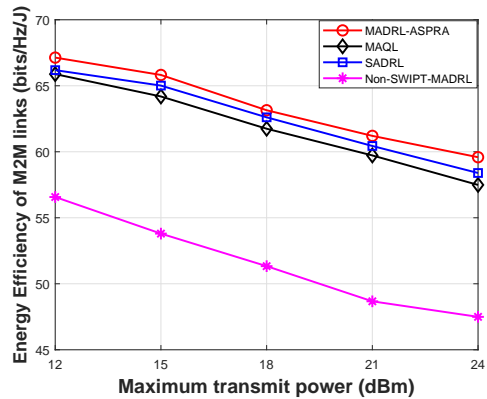}
\end{center}
\caption{Aggregate EE performance of M2M links with varying maximum transmit power.}
\label{fig:4}
\end{figure}
Fig.4 depicts the aggregate EE as a function of the maximum tolerable M2M transmit power $\overline{P}$. In this figure, it can be demonstrated clearly that the aggregate EE of M2M links decreases as the maximum transmit power threshold progressively rises. This is because although the large transmit power threshold can provide more available transmit power to each MTCD for payload delivery, more allocated power brings a higher level of power consumption. Furthermore, the aggregate EE of SWIPT schemes is higher due to the impact of EH. Additionally, the elicited results show that our proposed scheme outperforms other schemes under the same enabling conditions, and verify that the superiority of our scheme under different action set settings.
\begin{figure}[t]
\begin{center}
\includegraphics[width=.4\textwidth]{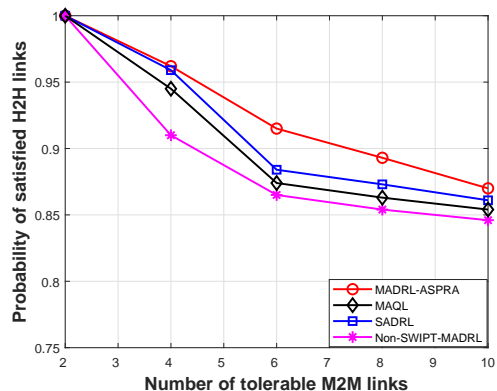}
\end{center}
\caption{Probability of satisfied H2H links with varying tolerable M2M links.}
\label{fig:5}
\end{figure}
\subsection{QoS Performance Comparisons}
Fig.5 describes the impact of the number of tolerable M2M links on the communication quality of H2H links. The degree of impact is expressed as the probability of satisfied H2H links, i.e., the ratio of the number of H2H links with guaranteed QoS requirements to the total number of H2H links.

From this figure, it can be shown that the probability of satisfied H2H links drops as the number of tolerable M2M links grows larger. The reason is that more interference will further reduces the received SINR of H2H links as a result of the increased number of sharing tolerable M2M links. Additionally, the satisfied probability of non-SWIPT scheme is particularly sensitive to the increase of tolerable M2M links when only a few tolerable M2M links exist to share spectrum with H2H links. Such rapid performance degradation comes later in SWIPT schemes, which can be obtained from the steep slope of the simulation curve. This phenomenon is caused by two main aspects. On one hand, in non-SWIPT scheme, tolerable M2M links give priority to sharing spectrum with H2H links rather than with critical M2M links. And due to the higher path loss, the tolerable M2M link will receive a weaker interference signal from the BS so that the tolerable M2M links can obtain higher EE. Meanwhile, non-SWIPT enabled critical M2M links can achieve higher EE without interference. On the other hand, in SWIPT schemes, tolerable M2M links give priority to sharing the spectrum of critical M2M links. This is due to the fact that application of SWIPT can significantly improve the EE when the suffered interference is controlled within an acceptable range. Moreover, we note that our proposed scheme always has the highest probability of satisfying H2H links along with the increase of shared M2M links. This is because that our scheme enables to obtain the best spectrum reuse selection for each tolerable M2M links through a distributed approach and efficient DNN approximators. The SADRL scheme is more inclined to assign spectrum sub-bands applicable to all agents, while the MAQL scheme possibly ignores some excellent selections.
\begin{figure}[t]
\begin{center}
\includegraphics[width=.4\textwidth]{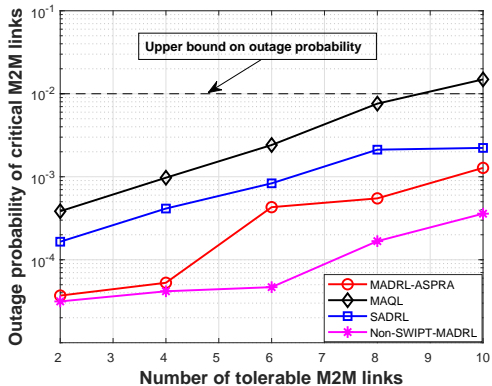}
\end{center}
\caption{Outage probability of critical M2M links with varying tolerable M2M links.}
\label{fig:6}
\end{figure}

Fig.6 demonstrates the reliability of critical M2M links versus the number of tolerable M2M links. It can be observed that the outage probability of critical M2M links increases with the growth of tolerable M2M links. This is based on the fact that the more spectrum access, the higher possibility of outage. Among these schemes, non-SWIPT scheme performs best. This is because non-SWIPT enabled critical M2M links dedicate all received energy to information decoding, resulting in high reliability. In contrast, SWIPT schemes convert the received interference into energy through EH to improve the EE, thus sacrificing certain reliability. In addition, the MAQL scheme cannot keep the outage probability of critical M2M links below targeted value $\overline{P_{s}}=0.01$ when the number of M2M links exceeds 8. The reason is attributed to the nature of RL: with an increased number of agents and enlarged state space, high computational complexity will yield ill-considered policies, thus leaving QoS requirements unguaranteed. Instead, the QoS requirements can be guaranteed by the DNN function approximator of DRL-based schemes, where multi-DNN outperforms single DNN. Moreover, the steep slope in the curve of our scheme corresponding to the situation that EE has barely fallen in Fig.3. The result indicates that the resource allocation strategy in our scheme increases the interference to critical M2M links in order to achieve higher EE performance.

Fig.7 presents the payload transmission probability of M2M payload transmission versus the number of tolerable M2M links. It can be appreciated that as the number of tolerable M2M links increases, the successful probability of payload transmission decreases accordingly. This occurs because excessive spectrum reuse also gives rise to interference and energy consumption, which reduces the capacity of each link to deliver packet data. Among all schemes, non-SWIPT scheme performs worst since it can only choose lower transmit power to maintain a higher EE performance, which leads to lower capacity. On the contrary, SWIPT schemes prefer to receive a certain amount of interference to achieve higher EE performance due to the EH function, thus increasing the chance of selecting larger transmit power for improving delivery rate. Furthermore, the proposed algorithm achieves the highest payload transmission probability since the multi-agent deployment are better able to fully consider the payload transmission requirements of each link than single-agent deployment. In contrast, the performance of MAQL is slightly poor, which is the inherent defect of the RL method as described above.
\begin{figure}[t]
\begin{center}
\includegraphics[width=.4\textwidth]{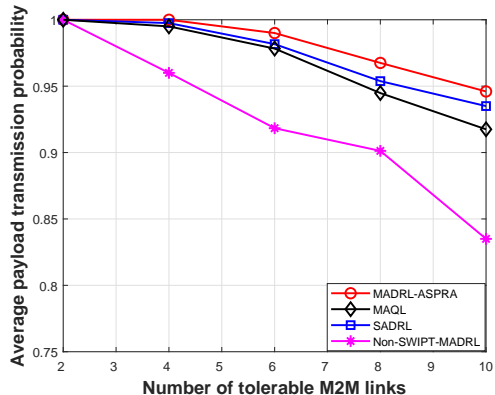}
\end{center}
\caption{Payload transmission probability of tolerable M2M links with varying tolerable M2M links.}
\label{fig:7}
\end{figure}
\section{CONCLUSION}\label{conclution}
This paper investigates the problem of energy-efficient resource allocation in cellular networks with H2H/M2M coexistence. In order to achieve harmonious and sustainable coexistence between the existing H2H and new M2M traffic, we first propose a SWIPT enabled network model in which both critical and tolerable MTCDs can simultaneously harvest energy and decode information. We then formulate the EE maximization problem subject to diverse QoS requirement constraints. Since the optimization problem is non-convex and the traditional techniques are hard to be applied, we further model the EE optimization of multiple M2M links as a multi-agent problem and develop an autonomous spectrum-power-ratio allocation scheme to obtain the global optimal solution. By using the designed behaviour-tracking based state and a common reward function, multiple agents can overcome the nonstationarity problem and ultimately pursue the global optimal policy under a cooperation mode. Finally, numerical results verify that compared with the existing schemes, the proposed scheme achieves higher EE performance under divers QoS requirements. In general, the proposed resource management scheme is of great significance for the future development of large-scale heterogeneous networks with H2H/M2M coexistence, and also provides some constructive ideas for theoretical researches and practical applications.

\bibliographystyle{IEEEtran}
\bibliography{mybib}
\begin{IEEEbiography}[{\includegraphics[width=1in,height=1.25in,clip,keepaspectratio]{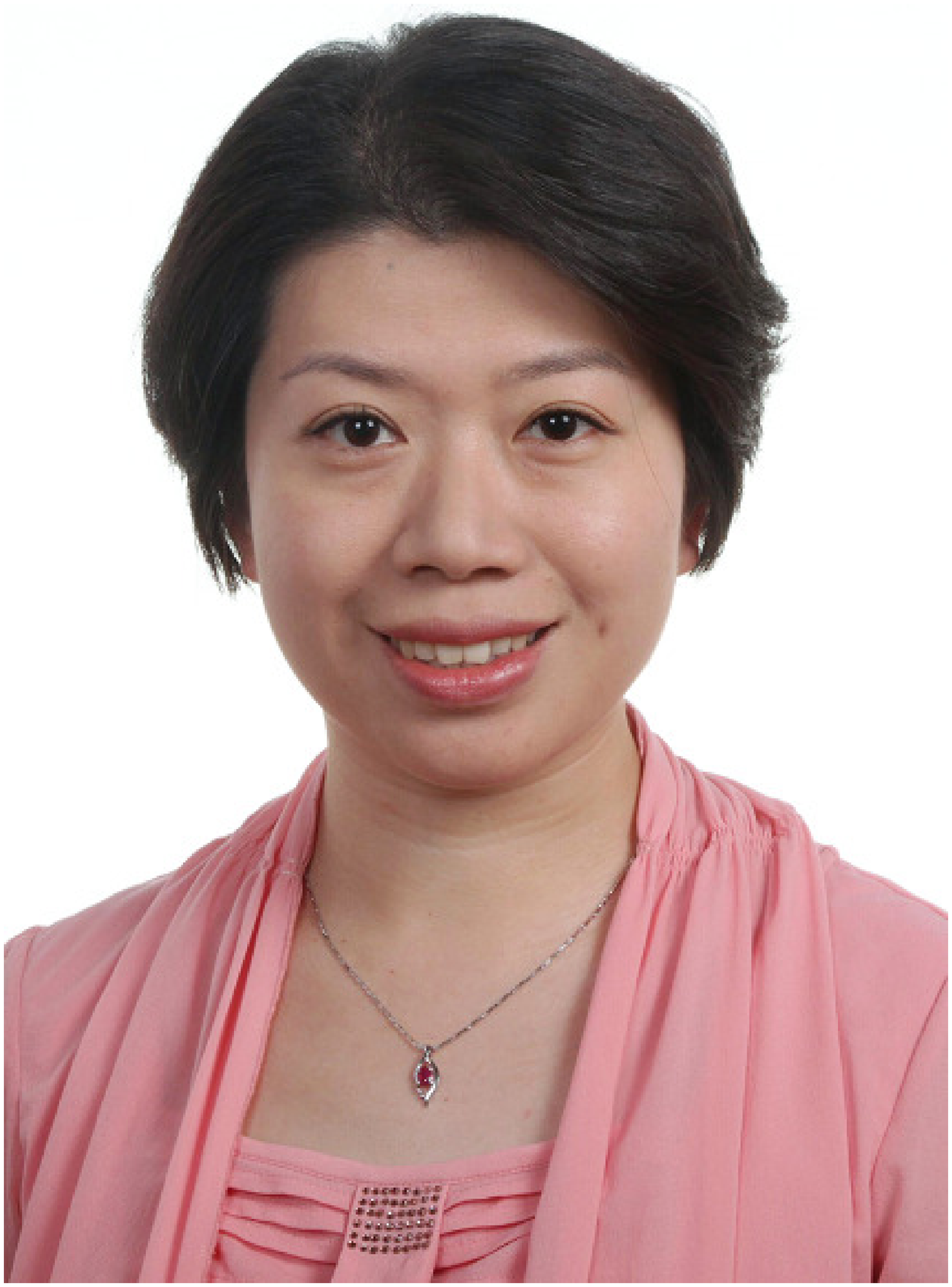}}]{Xuehua Li}
received the Ph.D. degree in telecommunications engineering from the Beijing University of Posts and Telecommunications, Beijing, China, in 2008. She is currently a Professor and the Deputy Dean of the School of Information and Communication Engineering with Beijing Information Science and Technology University, Beijing. She is a Senior Member of the Beijing Internet of Things Institute. Her research interests are in the broad areas of communications and information theory, particularly the Internet of Things, and coding for multimedia communications system.
\end{IEEEbiography}
\vspace{-1 cm}
\begin{IEEEbiography}[{\includegraphics[width=1in,height=1.25in,clip,keepaspectratio]{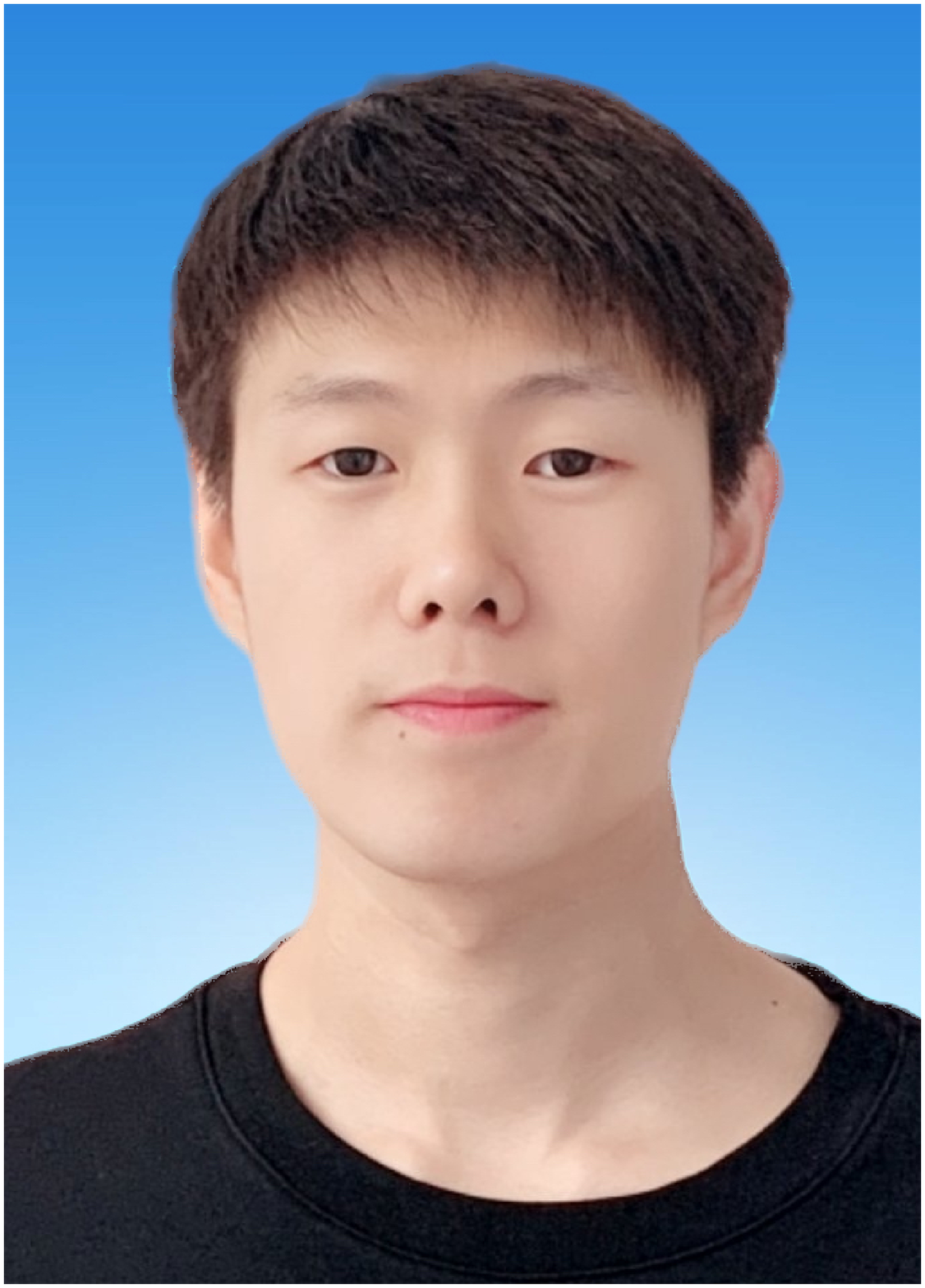}}]{Xing Wei}
is currently pursing the M.Phill degree with the School of Information and Communication Engineering, Beijing Information Science and Technology University. His research interests include wireless communications, machine-to-machine communications, radio resource management and machine learning.
\end{IEEEbiography}
\vspace{-1 cm}
\begin{IEEEbiography}[{\includegraphics[width=1in,height=1.25in,clip,keepaspectratio]{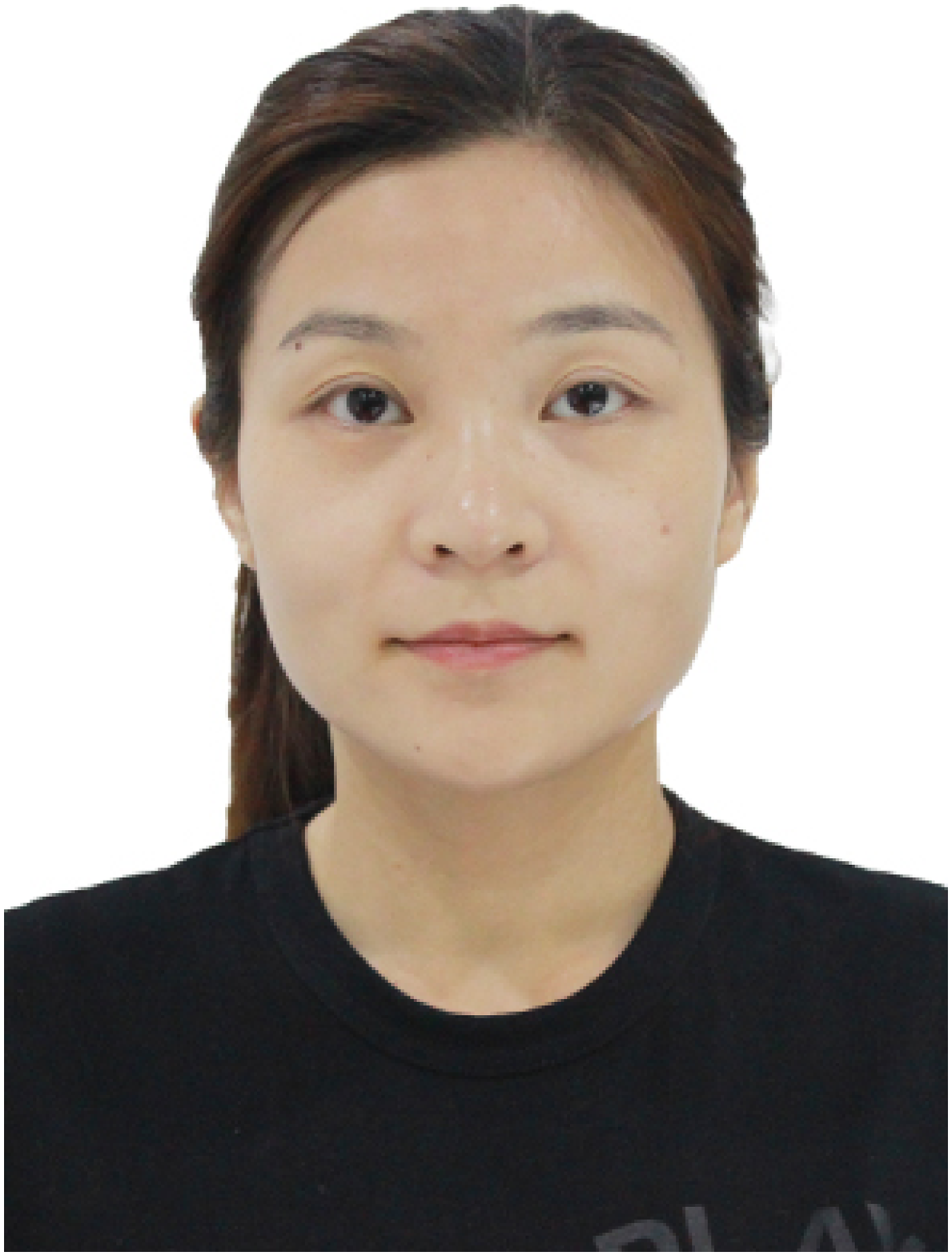}}]{Shuo Chen}
received the Ph.D. degree in Information and Communication Engineering from Beijing University of Posts and Telecommunication (BUPT) in 2018. She is currently an Associate Professor in the School of Information and Communication Engineering at Information and Communication Engineering (BISTU). Her current research interests are in the areas of wireless communications and networks, with an emphasis on resource management.
\end{IEEEbiography}
\vspace{-1 cm}
\begin{IEEEbiography}[{\includegraphics[width=1in,height=1.25in,clip,keepaspectratio]{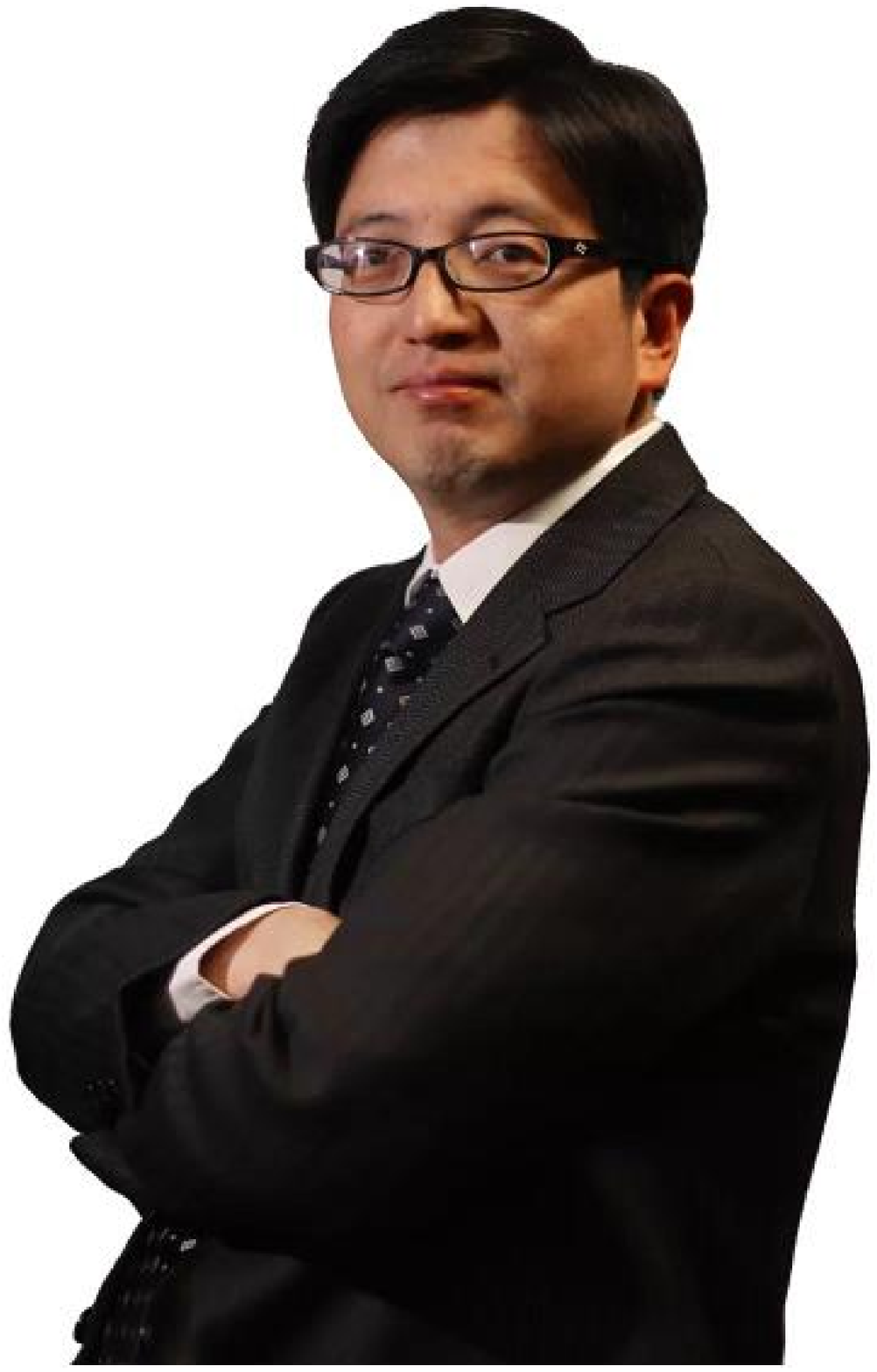}}]{Lixin Sun}
is the chairman and CEO of Baicells Technologies Co., Ltd, has spent more than 20 years in the mobile communications industry. He has held a number of positions and moderated standard meetings in driving the future Mobile technologies, including without limitation to: chairperson of ITU-R WP5D WG Technology and Vice, chair of ITU-R SG5, chair of APT AWG. Before he founded Baicells Technologies, he served as Huawei Fellow and the head of Strategy and Standard dept.
\end{IEEEbiography}
\end{document}